%% file: main.tex
\documentclass[reqno]{amsart}
\input{header.tex}
\input{preamble.tex}
\usepackage{hyperref}
\usepackage{mathrsfs}

\title[The Mathematical Structure of IIT]{The Mathematical Structure of Integrated Information Theory}

\author{Johannes Kleiner}
\email{johannes.kleiner@itp.uni-hannover.de}
\address{Munich Center for Mathematical Philosophy, University of Munich}

\author{Sean Tull}
\email{sean.tull@cambridgequantum.com}
\address{Cambridge Quantum Computing}

\begin{document}

\begin{abstract}
Integrated Information Theory is one of the leading models of consciousness. It aims to describe both the quality and quantity of the conscious experience of a physical system, such as the brain, in a particular state. In this contribution, we propound the mathematical structure of the theory, separating the essentials from auxiliary formal tools. We provide a definition of a generalized IIT which has IIT 3.0 of Tononi et.~al., as well as the Quantum IIT introduced by Zanardi et.~al. as special cases. This provides an axiomatic definition of the theory which may serve as the starting point for future formal investigations and as an introduction suitable for researchers with a formal background.
\end{abstract}

\maketitle

\input{1_intro}
\input{1_def}

\input{1_examples}

\input{1_probs}

\input{1_discussion}

\input{acknowledgements}
\bibliographystyle{alpha}
\bibliography{iit.bib}

\input{1_appendix}

\end{document}

%% file: header.tex
\usepackage{amsmath}
\usepackage{amsthm}
\usepackage{graphicx}
\usepackage{amssymb}
\usepackage{enumitem}

\usepackage[dvipsnames]{xcolor}

\theoremstyle{definition}
\newtheorem{Def}{Definition}
\newtheorem{definition}[Def]{Definition}

\newtheorem{Remark}[Def]{Remark}

\newtheorem{example}[Def]{Example}

\newcommand{\Thanks}{\vspace*{.5em} \noindent \thanks}
\newcommand{\intscaling}{\iota}

\newcommand{\la}{\langle}
\newcommand{\ra}{\rangle}

\newcommand{\R}{\mathbb{R}}

\DeclareMathOperator{\tr}{tr}

\renewcommand{\H}{\mathcal{H}}

\DeclareMathOperator{\norm}{| \hspace*{-0.1em}| }

\newcommand{\I}{{\mathcal{I}}}

\DeclareFontFamily{OT1}{rsfso}{}
\DeclareFontShape{OT1}{rsfso}{m}{n}{ <-7> rsfso5 <7-10> rsfso7 <10-> rsfso10}{}
\DeclareMathAlphabet{\mycal}{OT1}{rsfso}{m}{n}

\newcommand{\bei}{\begin{itemize}[itemsep=.2em]}
\newcommand{\eni}{\end{itemize}}

\usepackage{tikz,xypic}
\usetikzlibrary{decorations.pathreplacing,decorations.markings,arrows.meta,backgrounds,shapes}
\usetikzlibrary{circuits.ee.IEC}
\pgfdeclarelayer{edgelayer}
\pgfdeclarelayer{nodelayer}
\pgfsetlayers{background,edgelayer,nodelayer,main}
\tikzstyle{none}=[inner sep=0mm]
\tikzstyle{every loop}=[]
\tikzstyle{mark coordinate}=[inner sep=0pt,outer sep=0pt,minimum size=3pt,fill=black,circle]

\tikzset{arrow/.style={decoration={
    markings,
    mark=at position #1 with \arrow{>[length=2pt, width=3pt]}},
    postaction=decorate},
    reverse arrow/.style={decoration={
    markings,
    mark=at position #1 with {{\arrow{<[length=2pt, width=3pt]}}}},
    postaction=decorate}
}

\tikzstyle{upground}=[circuit ee IEC,thick,ground,rotate=90,scale=1.5]

\newcommand{\caus}{\mathsf{caus}}
\newcommand{\eff}{\mathsf{eff}}

\newcommand{\con}[2]{\mathbb{C}_{#1}(#2)}

\newcommand{\beq}{\begin{equation}}
\newcommand{\eeq}{\end{equation}}
\newcommand{\Proof}{\begin{proof}}
\newcommand{\QED}{\end{proof} \noindent}

%% file: preamble.tex
\usepackage{amsmath,amssymb,stmaryrd,xspace}
\usepackage{amsthm}
\usepackage{tikz-cd}
\usepackage{multirow}
\usepackage{mathtools}
\usepackage{url}
\usepackage{relsize}
\usepackage{bm} 
\newcounter{counter}

\newcommand{\nonext}[1]{#1'}

\newcommand{\DetSt}{\St}
\newcommand{\ProbSt}{\Prob}
\newcommand{\phimax}{\varphi^{\textrm{max}}}

\newcommand{\Expcat}{\cat{Exp}}

\newcommand{\PExp}{\mathbb{PE}}
\newcommand{\Exp}{\mathbb{E}}

\newcommand{\Sys}{\cat{Sys}}

\newcommand{\QShape}{\mathbb{Q}}
\newcommand{\cut}[2]{{#1}^{#2}}

\newcommand{\notetoself}[1]{}
\newcommand{\omitfornow}[1]{}
\newcommand{\omitthis}[1]{}

\newcommand{\Prob}{\mathcal P}

\newcommand{\Concepts}{\mathbb{C}}

\newcommand{\Split}[1]{\ensuremath{\mathrm{Split}}\xspace}

\newcommand{\cat}[1]{\ensuremath{\mathbf{#1}}\xspace}

\newcommand{\Sub}{\mathrm{Sub}}

\newcommand{\discardflip}[1]{\ensuremath{\tinygroundflipnew_{#1}}}

\newcommand{\uniform}[1]{{\omega_{#1}}} 

\DeclareFontFamily{U}{mathux}{\hyphenchar\font45}
\DeclareFontShape{U}{mathux}{m}{n}{
      <5> <6> <7> <8> <9> <10>
      <10.95> <12> <14.4> <17.28> <20.74> <24.88>
      mathux10
      }{}
\DeclareSymbolFont{mathux}{U}{mathux}{m}{n}

\DeclareMathSymbol{\bigovee}{1}{mathux}{"8F}

\DeclareMathSymbol{\bigperp}{1}{mathux}{"4E}

\DeclareFontFamily{U}{mathb}{\hyphenchar\font45}
\DeclareFontShape{U}{mathb}{m}{n}{
      <5> <6> <7> <8> <9> <10> gen * mathb
      <10.95> mathb10 <12> <14.4> <17.28> <20.74> <24.88> mathb12
      }{}
\DeclareSymbolFont{mathb}{U}{mathb}{m}{n}
\DeclareFontSubstitution{U}{mathb}{m}{n}
\DeclareMathSymbol{\mylgroup}{\mathbin}{mathb}{'160}
\DeclareMathSymbol{\myrgroup}{\mathbin}{mathb}{'161}

\newcommand{\St}{\text{St}}

\newcommand{\hilbH}{\mathcal{H}} 

\usepackage{tikz,xypic}

\usetikzlibrary{decorations.pathreplacing,decorations.markings,arrows.meta,backgrounds}
\pgfdeclarelayer{edgelayer}
\pgfdeclarelayer{nodelayer}
\pgfsetlayers{background,edgelayer,nodelayer,main}

\tikzstyle{cdot}=[circle, draw=black, fill=black!25, inner sep=.4ex] 
\tikzstyle{bigdot}=[dot, inner sep=0pt]
\tikzstyle{whitedot}=[circle, draw=black, fill=white, inner sep=.4ex]
\tikzstyle{greydot}=[circle, draw=black, fill=black!25, inner sep=.4ex] 
\tikzstyle{blackdot}=[circle, draw=black, fill=black, inner sep=.4ex]
\tikzset{arrow/.style={decoration={
    markings,
    mark=at position #1 with \arrow{>[length=2pt, width=3pt]}},
    postaction=decorate},
    reverse arrow/.style={decoration={
    markings,
    mark=at position #1 with {{\arrow{<[length=2pt, width=3pt]}}}},
    postaction=decorate}
}

\newenvironment{pic}[1][] {\begin{aligned}\begin{tikzpicture}[scale=2.0, font=\tiny,#1]}{\end{tikzpicture}\end{aligned}} 

\newif\ifvflip\pgfkeys{/tikz/vflip/.is if=vflip}
\newif\ifhflip\pgfkeys{/tikz/hflip/.is if=hflip}
\newif\ifhvflip\pgfkeys{/tikz/hvflip/.is if=hvflip}

\newenvironment{picc}[1][]
{\begin{aligned}\begin{tikzpicture}[font=\tiny,#1]}
{\end{tikzpicture}\end{aligned}}

\newlength\minimummorphismwidth
\newlength\stateheight
\newlength\minimumstatewidth
\newlength\connectheight
\tikzset{colour/.initial=white}

\tikzstyle{pure}=[line width=.7pt]

\makeatletter

\pgfdeclareshape{groundd}
{
    \savedanchor\centerpoint
    {
        \pgf@x=0pt
        \pgf@y=0pt
    }
    \anchor{center}{\centerpoint}
    \anchorborder{\centerpoint}

    \anchor{north}
    {
        \pgf@x=0pt
        \pgf@y=0.16\stateheight
    }
    \anchor{south}
    {
        \pgf@x=0pt
        \pgf@y=0pt
    }
    \saveddimen\overallwidth
    {
        \pgfkeysgetvalue{/pgf/minimum width}{\minwidth}
        \pgf@x=\minimumstatewidth
        \ifdim\pgf@x<\minwidth
            \pgf@x=\minwidth
        \fi
    }
    \backgroundpath
    {
        \begin{pgfonlayer}{main} 
        \pgfsetstrokecolor{black}
        \pgfsetlinewidth{1.25pt}
        \ifhflip
            \pgftransformyscale{-1}
        \fi
        \pgftransformscale{0.5}
        \pgfpathmoveto{\pgfpoint{-0.5*\overallwidth}{0pt}}
        \pgfpathlineto{\pgfpoint{0.5*\overallwidth}{0pt}}
        \pgfpathmoveto{\pgfpoint{-0.33*\overallwidth}{0.33*\stateheight}}
        \pgfpathlineto{\pgfpoint{0.33*\overallwidth}{0.33*\stateheight}}
        \pgfpathmoveto{\pgfpoint{-0.16*\overallwidth}{0.66*\stateheight}}
        \pgfpathlineto{\pgfpoint{0.16*\overallwidth}{0.66*\stateheight}}
        \pgfpathmoveto{\pgfpoint{-0.02*\overallwidth}{\stateheight}}
        \pgfpathlineto{\pgfpoint{0.02*\overallwidth}{\stateheight}}
        \pgfusepath{stroke}
        \end{pgfonlayer}
    }
}

\usepackage{tikz,xypic}
\usetikzlibrary{decorations.pathreplacing,decorations.markings,arrows.meta,backgrounds,shapes}
\usetikzlibrary{circuits.ee.IEC}
\pgfdeclarelayer{edgelayer}
\pgfdeclarelayer{nodelayer}
\pgfsetlayers{background,edgelayer,nodelayer,main}
\tikzstyle{none}=[inner sep=0mm]
\tikzstyle{every loop}=[]
\tikzstyle{mark coordinate}=[inner sep=0pt,outer sep=0pt,minimum size=3pt,fill=black,circle]

\tikzset{arrow/.style={decoration={
    markings,
    mark=at position #1 with \arrow{>[length=2pt, width=3pt]}},
    postaction=decorate},
    reverse arrow/.style={decoration={
    markings,
    mark=at position #1 with {{\arrow{<[length=2pt, width=3pt]}}}},
    postaction=decorate}
}

\tikzstyle{upground}=[circuit ee IEC,thick,ground,rotate=90,scale=1.5]
\tikzstyle{upgroundwhite}=[circuit ee IEC,thick,ground,rotate=90,scale=1.5, fill=white]
\tikzstyle{downground}=[circuit ee IEC,thick,ground,rotate=-90,scale=1.5]
\tikzstyle{downgroundnorm}=[circuit ee IEC,thick,ground,rotate=-90,scale=1.5, fill=white]

\newcommand{\mapminh}{5mm} 
\newcommand{\stateminh}{5mm}
\newcommand{\maplw}{0.7pt} 
\newcommand{\stateshift}{-0.2pt}
\newcommand{\effectshift}{-0.2pt}

\tikzstyle{box}=[map]
\tikzstyle{medium box}=[medium map]
\tikzstyle{dot}=[inner sep=0mm,minimum width=2mm,minimum height=2mm,draw,shape=circle]  
\tikzstyle{black dot}=[dot,fill=black]
\tikzstyle{white dot}=[dot,fill=white,,text depth=-0.2mm]
\tikzstyle{grey dot}=[dot,fill=black!25] 

\tikzstyle{corner1}=[box,fill=white, font=\footnotesize] %
\tikzstyle{corner2}=[dot,fill=white, font=\footnotesize] %
\tikzstyle{corner3}=[dot,fill=black!25, font=\footnotesize] %
\tikzstyle{corner4}=[dot,fill=black, font=\footnotesize] %

\tikzstyle{scalar}=[circle,draw,inner sep=2pt, line width=\maplw] 

\usetikzlibrary{shapes.misc, positioning}

\tikzset{stateshape/.style={append after command={
   \pgfextra
        \draw[sharp corners, fill=white, line width = \maplw]%
    (\tikzlastnode.west)%
    [rounded corners=0pt] |- (\tikzlastnode.north)%
    [rounded corners=0pt] -| (\tikzlastnode.east)%
    [rounded corners=5pt] |- (\tikzlastnode.south)%
    [rounded corners=5pt] -| (\tikzlastnode.west);
   \endpgfextra}}}

\tikzset{effectshape/.style={append after command={
   \pgfextra
        \draw[sharp corners, fill=white, line width = \maplw]%
    (\tikzlastnode.west)%
    [rounded corners=0pt] |- (\tikzlastnode.south)%
    [rounded corners=0pt] -| (\tikzlastnode.east)%
    [rounded corners=5pt] |- (\tikzlastnode.north)%
    [rounded corners=5pt] -| (\tikzlastnode.west);
   \endpgfextra}}}

 \tikzstyle{map}=[draw,shape=rectangle, inner sep=2pt,minimum height=\mapminh, minimum width=7mm,fill=white]

\tikzstyle{point}=[stateshape,inner sep=2pt, minimum width=6mm, minimum height=\stateminh, yshift=\stateshift]
\tikzstyle{copoint}=[effectshape,inner sep=.2pt, minimum width=6mm, minimum height=\stateminh, yshift=-\effectshift]

\tikzstyle{wide point}=[point, minimum width=12mm]
\tikzstyle{wide copoint}=[copoint, minimum width=12mm]

\tikzstyle{decomp}=[fill=white,draw,shape=isosceles triangle,shape border rotate=-90,isosceles triangle stretches=true,inner sep=0pt,minimum width=0.75cm,minimum height=4mm,yshift=-0.0mm]

\tikzstyle{decompwide}=[fill=white,draw,shape=isosceles triangle,shape border rotate=-90,isosceles triangle stretches=true,inner sep=0pt,minimum width=1.5cm,minimum height=4mm,yshift=-0.0mm]

\tikzstyle{decompflip}=[fill=white,draw,shape=isosceles triangle,shape border rotate=90,isosceles triangle stretches=true,inner sep=0pt,minimum width=0.75cm,minimum height=4mm,yshift=-0.0mm]

\tikzstyle{decompwideflip}=[fill=white,draw,shape=isosceles triangle,shape border rotate=90,isosceles triangle stretches=true,inner sep=0pt,minimum width=1.5cm,minimum height=4mm,yshift=-0.0mm]

 \tikzstyle{map}=[draw,shape=rectangle, inner sep=2pt,minimum height=\mapminh, minimum width=7mm,fill=white, line width = \maplw]

\tikzstyle{medium map} = [map, minimum width = 12mm] 
\tikzstyle{semilarge map} = [map, minimum width = 15mm] 
\tikzstyle{large map} = [map, minimum width = 18mm] 

\tikzstyle{kpoint} =[point]
\tikzstyle{kpointadj} =[copoint]
\tikzstyle{kpointconj}=[dagpointconj] 

\makeatletter
\newcommand{\boxshape}[3]{%
\pgfdeclareshape{#1}{
\inheritsavedanchors[from=rectangle] 
\inheritanchorborder[from=rectangle]
\inheritanchor[from=rectangle]{center}
\inheritanchor[from=rectangle]{north}
\inheritanchor[from=rectangle]{south}
\inheritanchor[from=rectangle]{west}
\inheritanchor[from=rectangle]{east}
\backgroundpath{
\southwest \pgf@xa=\pgf@x \pgf@ya=\pgf@y
\northeast \pgf@xb=\pgf@x \pgf@yb=\pgf@y

\@tempdima=#2
\@tempdimb=#3

\pgfpathmoveto{\pgfpoint{\pgf@xa - 5pt + \@tempdima}{\pgf@ya}}
\pgfpathlineto{\pgfpoint{\pgf@xa - 5pt - \@tempdima}{\pgf@yb}}
\pgfpathlineto{\pgfpoint{\pgf@xb + 5pt + \@tempdimb}{\pgf@yb}}
\pgfpathlineto{\pgfpoint{\pgf@xb + 5pt - \@tempdimb}{\pgf@ya}}
\pgfpathlineto{\pgfpoint{\pgf@xa - 5pt + \@tempdima}{\pgf@ya}}
\pgfpathclose
}
}}

\boxshape{NEbox}{0pt}{3pt} 
\boxshape{SEbox}{0pt}{-3pt}
\boxshape{NWbox}{3pt}{0pt}
\boxshape{SWbox}{-3pt}{0pt}
\boxshape{rec-box}{0pt}{0pt}
\makeatother

\tikzstyle{cloud}=[shape=cloud,draw,minimum width=1.5cm,minimum height=1.5cm]

\tikzstyle{dagmap}=[draw,shape=NEbox,inner sep=2pt,minimum height=\mapminh,fill=white, line width = \maplw] %
\tikzstyle{dashedmap}=[draw,dashed,shape=NEbox,inner sep=2pt,minimum height=\mapminh,fill=white, line width = \maplw]
\tikzstyle{mapdag}=[draw,shape=SEbox,inner sep=2pt,minimum height=\mapminh,fill=white, line width = \maplw]
\tikzstyle{mapadj}=[draw,shape=SEbox,inner sep=2pt,minimum height=\mapminh,fill=white, line width = \maplw]
\tikzstyle{maptrans}=[draw,shape=SWbox,inner sep=2pt,minimum height=\mapminh,fill=white, line width = \maplw]
\tikzstyle{mapconj}=[draw,shape=NWbox,inner sep=2pt,minimum height=\mapminh,fill=white, line width = \maplw]

\tikzstyle{medium dagmap}=[draw,shape=NEbox,inner sep=2pt,minimum height=\mapminh,fill=white,minimum width=7mm, line width = \maplw]
\tikzstyle{semilarge dagmap}=[draw,shape=NEbox,inner sep=2pt,minimum height=\mapminh,fill=white,minimum width=9.5mm, line width = \maplw]
\tikzstyle{large dagmap}=[draw,shape=NEbox,inner sep=2pt,minimum height=\mapminh,fill=white,minimum width=12mm, line width = \maplw]

\makeatletter

\pgfdeclareshape{cornerpoint}{
\inheritsavedanchors[from=rectangle] 
\inheritanchorborder[from=rectangle]
\inheritanchor[from=rectangle]{center}
\inheritanchor[from=rectangle]{north}
\inheritanchor[from=rectangle]{south}
\inheritanchor[from=rectangle]{west}
\inheritanchor[from=rectangle]{east}
\backgroundpath{
\southwest \pgf@xa=\pgf@x \pgf@ya=\pgf@y
\northeast \pgf@xb=\pgf@x \pgf@yb=\pgf@y

\pgfmathsetmacro{\pgf@shorten@left}{\pgfkeysvalueof{/tikz/shorten left}}
\pgfmathsetmacro{\pgf@shorten@right}{\pgfkeysvalueof{/tikz/shorten right}}

\pgfpathmoveto{\pgfpoint{0.5 * (\pgf@xa + \pgf@xb)}{\pgf@ya - 5pt}}
\pgfpathlineto{\pgfpoint{\pgf@xa - 8pt + \pgf@shorten@left}{\pgf@yb - 1.5 * \pgf@shorten@left}}
\pgfpathlineto{\pgfpoint{\pgf@xa - 8pt + \pgf@shorten@left}{\pgf@yb}}
\pgfpathlineto{\pgfpoint{\pgf@xb + 8pt - \pgf@shorten@right}{\pgf@yb}}
\pgfpathlineto{\pgfpoint{\pgf@xb + 8pt - \pgf@shorten@right}{\pgf@yb - 1.5 * \pgf@shorten@right}}
\pgfpathclose
}
}

\pgfdeclareshape{cornercopoint}{
\inheritsavedanchors[from=rectangle] 
\inheritanchorborder[from=rectangle]
\inheritanchor[from=rectangle]{center}
\inheritanchor[from=rectangle]{north}
\inheritanchor[from=rectangle]{south}
\inheritanchor[from=rectangle]{west}
\inheritanchor[from=rectangle]{east}
\backgroundpath{
\southwest \pgf@xa=\pgf@x \pgf@ya=\pgf@y
\northeast \pgf@xb=\pgf@x \pgf@yb=\pgf@y

\pgfmathsetmacro{\pgf@shorten@left}{\pgfkeysvalueof{/tikz/shorten left}}
\pgfmathsetmacro{\pgf@shorten@right}{\pgfkeysvalueof{/tikz/shorten right}}

\pgfpathmoveto{\pgfpoint{0.5 * (\pgf@xa + \pgf@xb)}{\pgf@yb + 5pt}}
\pgfpathlineto{\pgfpoint{\pgf@xa - 8pt + \pgf@shorten@left}{\pgf@ya + 1.5 * \pgf@shorten@left}}
\pgfpathlineto{\pgfpoint{\pgf@xa - 8pt + \pgf@shorten@left}{\pgf@ya}}
\pgfpathlineto{\pgfpoint{\pgf@xb + 8pt - \pgf@shorten@right}{\pgf@ya}}
\pgfpathlineto{\pgfpoint{\pgf@xb + 8pt - \pgf@shorten@right}{\pgf@ya + 1.5 * \pgf@shorten@right}}
\pgfpathclose
}
}

\makeatother

\pgfkeyssetvalue{/tikz/shorten left}{0pt}
\pgfkeyssetvalue{/tikz/shorten right}{0pt}

\tikzstyle{dagpoint common}=[draw,fill=white,inner sep=1pt, line width = \maplw, minimum height = 4mm, yshift=1.2pt] 
\tikzstyle{dagpoint sc}=[shape=cornerpoint,dagpoint common]
\tikzstyle{dagpoint adjoint sc}=[shape=cornercopoint,dagpoint common]
\tikzstyle{dagpoint}=[shape=cornerpoint,shorten left=4pt,dagpoint common]
\tikzstyle{dagpointadj}=[shape=cornercopoint,shorten left=5pt,dagpoint common]
\tikzstyle{dagpointconj}=[shape=cornerpoint,shorten right=5pt,dagpoint common]
\tikzstyle{dagpointtrans}=[shape=cornercopoint,shorten right=5pt,dagpoint common]
\tikzstyle{dagpointsymm}=[shape=cornerpoint,shorten left=5pt,shorten right=5pt,dagpoint common]

\tikzstyle{widedagpoint}=[dagpoint, minimum width=1 cm, inner sep=2pt]
\tikzstyle{widedagpointadj}=[dagpointadj, minimum width=1 cm, inner sep=2pt]

\tikzstyle{every picture}=[baseline=-0.25em,scale=0.5]
\tikzstyle{label}=[font=\footnotesize,text height=1ex, text depth=0.15ex]

\usetikzlibrary[shapes]

\tikzset{
sidetriangle/.style = {regular polygon, regular polygon sides = 3, aspect = 1, shape border rotate = 90, draw, inner sep = 0, minimum width = 1.2cm}
}

\tikzset{
isoc/.style = {shape=isosceles triangle, shape border rotate = 180, isosceles triangle stretches = true, minimum width = 1.2cm, minimum height= 1.5cm, inner sep = 0.3}}

\tikzset{
coarse/.style = {shape = circle, fill = white, draw, inner sep = 0, minimum width =0.125cm}
}
\tikzset{
coarsesymbol/.style = {shape = circle, fill = white, inner sep = -0.7, minimum width = 0.125cm}
}

\tikzstyle{sidetriangle2}=[sidetriangle, minimum width = 2cm, fill=white]
\tikzstyle{sideisocsmall}]=[style=isoc, minimum width = 1cm, minimum height = 0.8cm, draw, fill=white, font=\Large]
\tikzstyle{sideisoc}]=[style=isoc, minimum width = 2cm, draw, fill=white, font=\Large]
\tikzstyle{sideisocmid}]=[style=isoc, minimum width = 2.5cm, draw, fill=white, font=\Large]
\tikzstyle{sideisocmedium}]=[style=isoc, minimum width = 3cm, draw, fill=white, font=\Large]

\newcommand{\tinygroundflipnew}{
\smash{
{\hspace{-3pt}
\ensuremath{
\begin{picc}[yscale=-1.0] 
    \node[upground, xscale=0.8, yscale=-0.7] (1) at (0,0.10) {};
    \draw (0,-0.03) to (0,-0.31);
\end{picc}
}\hspace{-1pt}}}}

\newcommand{\Cond}[1]{{\scriptstyle |}#1{\scriptstyle \ra}}
\newcommand{\Mar}[1]{{\scriptstyle \la #1 |}}
\newcommand{\Bot}{{\scriptscriptstyle \bot}}
\newcommand{\Botl}{{\!\scriptscriptstyle \bot}}

%% file: 1_intro.tex
\section{Introduction}

\emph{Integrated Information Theory (IIT)}, developed by Giulio Tononi and collaborators, has emerged as one of the leading scientific theories of consciousness~\cite{oizumi2014phenomenology,marshall2016integrated,tononi2016integrated,mayner2018pyphi,koch2016neural}. At the heart of the theory is an algorithm which, based on the level of \emph{integration} of the internal functional relationships of a physical system in a given state, aims to determine both the quality and quantity (`$\Phi$ value') of its conscious experience.

While promising in itself, the mathematical formulation of the theory is not satisfying to date. The presentation in terms of examples and concomitant explanation veils the essential mathematical structure of the theory and impedes philosophical and scientific analysis. In addition, the current definition of the theory can only be applied to quite simple classical physical systems~\cite{Barrett.2014}, which is problematic if the theory is taken to be a fundamental theory of consciousness, and should eventually be reconciled with our present theories of physics.

To resolve these problems, we examine the essentials of the IIT algorithm and formally define a generalized notion of Integrated Information Theory.
This notion captures the inherent mathematical structure of IIT and offers a rigorous mathematical definition of the theory which has `classical' IIT 3.0 of Tononi et. al.~\cite{oizumi2014phenomenology, marshall2016integrated,mayner2018pyphi} as well as the more recently introduced \emph{Quantum Integrated Information Theory} of Zanardi, Tomka and Venuti~\cite{zanardi2018quantum} as special cases. In addition, this generalization allows us to extend classical IIT, freeing it from a number of simplifying assumptions identified in~\cite{BarrettMediano.2019}.

In the associated article~\cite{IITProcess} we show more generally how the main notions of IIT, including causation and integration, can be treated, and an IIT defined, starting from any suitable theory of physical systems and processes described in terms of category theory. Restricting to classical or quantum process then yields each of the above as special cases. This treatment makes IIT applicable to a large class of physical systems and helps overcome the current restrictions.

Our definition of IIT may serve as the starting point for further mathematical analysis of IIT, in particular if related to category theory~\cite{tsuchiya2016using,northoff2019mathematics}. It also provides a simplification and mathematical clarification of the IIT algorithm which 
extends the technical analysis of the theory \cite{Barrett.2014,tegmark2015consciousness,tegmark2016improved} and may
contribute to its ongoing critical discussion \cite{Bayne.2018,mediano2019measuring,mediano2019beyond,tononi2015consciousness}. The concise presentation of IIT in this article should also help to make IIT more easily accessible for mathematicians, physicists and other researchers with a strongly formal background.

\begin{figure}
\[
\scalebox{0.9}{
\begin{tikzpicture}[node distance = 18em, auto]
\tikzstyle{block} = [rectangle, draw, text centered, rounded corners]
\node [block, text width = 8em] (phys) {Physical systems and states};
\node [block, right of=phys, text width = 10em] (ment) {Spaces and states of \\conscious experience};
\draw [->, line width = 0.1em] (phys) edge (ment);
\node [style=label] at (6, .4) {$\Exp$};
\end{tikzpicture}
}
\]
\caption{\small An Integrated Information Theory specifies for every system in a particular state its conscious experience, described formally as an element of an experience space. In our formalization, this is a map
\[ 
\! \! \! \! \! \! \! \! \! \! \! \! \! \! \! \! \!
\! \! \! \! \! \! \! \! \! \! \! \! \! \! \! \! \!
\! \! \! \! \! \! \! \! 
\Sys
\stackrel{\Exp}{\longrightarrow}  \Expcat 
\]
from the system class $\Sys$ into a class $\Expcat$ of experience spaces, which, first,
sends each system $S$ to its space of possible experiences $\Exp(S)$, and, second,
sends each state $s \in \St(S)$ to the actual experience the system is having when in that space,
\[
\! \! \! \! \! \! \! \! \! \! \! \! \! \! \! \! \!
\! \! \! \! \! \! \! \! \! \! \! \! \! \! \! \! \!
\St(S) \to \Exp(S) \qquad 
s \mapsto \Exp(S,s) \:.
\]
The definition of this map in terms of axiomatic descriptions of physical systems, experience spaces and further structure used in classical IIT is given in the first half of this paper.
}\label{FigureBox}
\end{figure}

\subsection{Structure of article}
We begin by introducing the necessary ingredients of a generalised Integrated Information Theory in Sections~\ref{sec:sys} to~\ref{sec:repertoires}, 
namely physical systems, spaces of conscious experience and cause-effect repertoires. Our approach is \emph{axiomatic} in that we state only the precise formal structure which is necessary to apply the IIT algorithm. In Section~\ref{sec:Integration}, we introduce a simple formal tool which allows us to present the definition of the algorithm of an IIT in a concise form in Sections~\ref{sec:alg-mech} and~\ref{sec:alg-sys}. Finally, in Section~\ref{sec:IITs}, we summarise the full definition of such a theory.

Following this we give several examples including IIT 3.0 in Section~\ref{sec:Classical} and Quantum IIT in Section~\ref{sec:quantum-IIT}. In Section~\ref{sec:problems} we discuss how our formulation allows one to extend classical IIT in several fundamental ways, before discussing further modifications to our approach and other future work in Section~\ref{sec:discussion}.  Finally, the appendix includes a detailed explanation of how our generalization of IIT coincides with its usual presentation in the case of classical IIT. 

%% file: 1_def.tex
\section{Systems} \label{sec:sys}

The first step in defining an Integrated Information Theory (IIT) is to specify a class $\Sys$ of physical \emph{systems} to be studied. Each element $S \in \Sys$ is interpreted as a model of one particular physical system. In order to apply the IIT algorithm, it is only necessary that each element $S$ come with the following features.

\begin{definition} \label{def:sys-class}
A \emph{system class} $\Sys$ is a class each of whose elements $S$, called \emph{systems}, come with the following data:
\begin{enumerate}[label=\arabic*.]
\item a set $\St(S)$ of \emph{states};
\item for every $s \in \St(S)$, a set $\Sub_s(S) \subset \Sys$ of subsystems and for each $M \in \Sub_s(S)$ an
induced state $s|_M \in \St(M)$;
\item \label{it:Decomp} a set $\mathbb D_S$ of {\em decompositions}, with a given \emph{trivial decomposition} $1 \in \mathbb D_S$;
\item for each $z \in \mathbb D_S$ a corresponding \emph{cut system} $\cut{S}{z} \in \Sys$ and for each state $s \in \St(S)$ a corresponding \emph{cut state} $\cut{s}{z} \in \St(\cut{S}{z})$.
\end{enumerate}
\end{definition}
Moreover, we require that $\Sys$ contains a distinguished \emph{empty system}, denoted~$I$, and that $I \in \Sub(S)$ for all $S$. For the IIT algorithm to work, we need to assume furthermore that the number of subsystems remains the same under cuts or changes of states,
i.e. $\Sub_s(S) \simeq \Sub_{s'}(S)$ for all $s, s' \in \St(S)$ and $\Sub_s(S) \simeq \Sub_{\cut{s}{z}}(\cut{S}{z})$ for all $z \in \mathbb D_S$. Here, $\simeq$ indicates bijections. Note that subsystems of a system may depend on the state of the system, in accordance with classical IIT. 

In this article we will assume that each set $\Sub_s(S)$ is finite, discussing the extension to the infinite case in Section \ref{sec:discussion}. We will give examples of system classes and for all following definitions in Sections~\ref{sec:Classical} and~\ref{sec:quantum-IIT}.

\section{Experience} \label{sec:Experience}

An IIT aims to specify for each system in a particular state its \emph{conscious experience}. As such, it will require a mathematical model of such experiences. Examining classical IIT, we find the following basic features of the final experiential states it describes which are needed for its algorithm.

Firstly, each experience $e$ should crucially come with an \emph{intensity}, given by a number $\norm e \norm$ in the non-negative reals  $\R^+$ (including zero). 
This intensity will finally correspond to the overall intensity of experience, usually denoted by $\Phi$. Next, in order to compare experiences, we require a notion of \emph{distance} $d(e,e')$ between any pair of experiences $e, e'$. Finally, the algorithm will require us to be able to \emph{rescale} any given experience $e$ to have any given intensity. Mathematically, this is most easily encoded by letting us multiply any experience $e$ by any number $r \in \R^+$.
In summary, a minimal model of experience in a generalised IIT is the following.

\begin{definition} \label{def:exp-space}
An \emph{experience space} is a set $E$ with: 
\begin{enumerate}[label=\arabic*.]
\item 
an \emph{intensity} function $\norm . \norm \colon E \to \R^+$;
\item 
a \emph{distance} function $d \colon E \times E \to \mathbb{R}^+$;
\item 
a \emph{scalar multiplication} $\R^+ \times E \to E$, denoted $(r,e) \mapsto r \cdot e$, satisfying
\begin{align*}
\norm r \cdot e \norm = r \cdot \norm e \norm \qquad 
r \cdot (s \cdot e) = (rs) \cdot e \qquad
1 \cdot e = e
\end{align*}
for all $e \in E$ and $r, s \in \R^+$.
\end{enumerate}
\end{definition}

We remark that this same axiomatisation will apply both to the full space of experiences of a system, as well as to the the spaces describing components of the experiences (`concepts' and `proto-experiences' defined in later sections). We note that the distance function does not necessarily have to satisfy the axioms of a metric. While this and further natural axioms such as $d(r \cdot e, r \cdot f) = r \cdot d(e,f)$ might hold, they are not necessary for the IIT algorithm.

The above definition is very general, and in any specific theory the experiences may come with richer further structure. The following example describes the experience space used in classical IIT.

\begin{example} \label{ex:IITProto}
Any metric space $(X,d)$ may be extended to an experience space $\bar X := X \times \mathbb{R}^+$ in various ways. E.g., 
one can define $\norm (x,r) \norm = r$, $r \cdot (x,s) = (x, rs)$ and define the distance as
\begin{equation}\label{eq:ExIITProto}
d\big((x,r),(y,s)\big) = r \, d(x,y) 
\end{equation}
This is the definition used in classical IIT (cf. Section~\ref{sec:Classical} and Appendix~\ref{sec:comparison}).
\end{example}

An important operation on experience spaces is taking their \emph{product}. 

\begin{definition} \label{def:prod-spaces}
For experience spaces $E$ and $F$, we define the product to be the space $E \times F$ with distance
\begin{equation}\label{eq:DistanceProduct}
d\big( (e,f) , (e',f') \big) = d(e,e') + d(f, f')  \:,
\end{equation} 
intensity $\norm (e,f) \norm = \max \{ \norm e \norm, \norm f \norm \}$ and scalar multiplication $r \cdot (e,f) = (r \cdot e, r \cdot f)$.
This generalises to any finite product $\prod_{i \in I} E_i$ of experience spaces.
\end{definition}

\section{Repertoires} \label{sec:repertoires}


In order to define the experience space and individual experiences of a system $S$, an IIT utilizes basic building blocks called `repertoires', which we will now define.
Next to the specification of a system class, this is the essential data necessary for the IIT algorithm to be applied.

Each repertoire describes a way of `decomposing' experiences, in the following sense. Let $D$ denote any set with a distinguished element $1$, for example the set $\mathbb D_S$ of decompositions of a system $S$, where the distinguished element is the trivial decomposition $1 \in \mathbb D_S$.

\begin{definition}\label{def:Decomp}
Let $e$ be an element of an experience space $E$. A \emph{decomposition of $e$ over $D$} is a mapping $\bar{e} \colon D \to E$ with $\bar{e}(1) = e$.
\end{definition}

In more detail, a repertoire specifies a proto-experience for every pair of subsystems and describes how this experience changes if the subsystems are decomposed.
This allows one to assess how integrated the system is with respect to a particular repertoire. Two repertoires are necessary for the IIT algorithm to be applied, together called the cause-effect repertoire.

For subsystems $M,P \in \Sub_s(S)$, define $\mathbb D_{M,P} := \mathbb D_M \times \mathbb D_P$. This set describes the decomposition of both subsystems simultaneously. It has a distinguished element $1 = (1_M, 1_P)$.

\begin{definition} \label{def:repertoire}
A \emph{cause-effect repertoire} at $S$ is given by a choice of experience space $\PExp(S)$, called the space of \emph{proto-experiences}, and for each 
$s \in \St(S)$ and $M, P \in \Sub_s(S)$, a pair of elements 
\begin{equation}\label{eq:CauseEffectRep}
\caus_s(M,P) \  ,\  \eff_s(M,P) \  \in \ \PExp(S)
\end{equation}
and for each of them a decomposition over $\mathbb D_{M,P}$.
\end{definition}

Examples of cause-effect repertoires will be given in Sections~\ref{sec:Classical} and~\ref{sec:quantum-IIT}. A general definition in terms of process theories is given in~\cite{IITProcess}. For the IIT algorithm, a cause-effect repertoire needs to be specified for every system $S$, as in the following definition.

\begin{definition}\label{def:cestructure}
A \emph{cause-effect structure} is a specification of a cause-effect repertoire for every $S \in \Sys$ such that
\begin{equation}\label{eq:cestructure}
\PExp(S) = \PExp(\cut{S}{z}) \quad \textrm{ for all } \quad z \in \mathbb D_S \:.
\end{equation}
\end{definition}

The names `cause' and `effect' highlight that the definitions of $\caus_s(M,P)$ and  $\eff_s(M,P)$ in classical and quantum IIT describe the causal dynamics of the system. More precisely, they are intended to capture the manner in which the `current' state~$s$ of the system, when restricted to $M$, constrains the `previous' or `next' state of $P$, respectively.

\section{Integration} \label{sec:Integration}

We have now introduced all of the data required to define an IIT; namely, a system class along with a cause-effect structure. From this, we will give an algorithm aiming to specify the conscious experience of a system. Before proceeding to do so, we introduce a conceptual short-cut
which allows the algorithm to be stated in a concise form. This captures the core ingredient of an IIT, namely the computation of how integrated an entity is.

\begin{definition}\label{def:IntScaling}
Let $E$ be an experience space and $e$ an element with a decomposition over some set $D$. The \emph{integration level} of $e$ relative to this decomposition is 
\begin{equation}\label{eq:IntegrLevel}
\phi(e) := \min_{1 \neq z \in D} d(e, \bar{e}(z)) \:.
\end{equation}
Here, $d$ denotes the distance function of $E$, and the minimum is taken over all elements of $D$ besides $1$. The \emph{integration scaling} of $e$ is then the element of $E$ defined by
\begin{equation}\label{eq:IntegrData}
	\intscaling(e) := \phi(e) \cdot \hat{e} \:,
\end{equation}
where $\hat e$ denotes the \emph{normalization} of $e$, defined as
\begin{align*}
\hat e := \begin{cases}
\frac{1}{\norm e \norm} \cdot e & \textrm{ if } \norm e \norm \neq 0 \\
 e & \textrm{ if } \norm e \norm = 0 \:.
\end{cases}
\end{align*}
Finally, the \emph{integration scaling} of a pair $e_1, e_2$ of such elements is the pair
\begin{equation}\label{eq:IntScalingPair}
\intscaling(e_1, e_2) :=
(\phi \cdot \hat{e_1},
\phi \cdot \hat{e_2})
\end{equation}
where $\phi := \min(\phi(e_1), \phi(e_2))$
is the minimum of their integration levels.
\end{definition}

We will also need to consider indexed collections of decomposable elements.
Let $S$ be a system in a state $s \in \St(S)$ and assume that for every $M \in \Sub_s(S)$
an element $e_M$ of some experience space $E_M$ with a decomposition over some $D_M$ is given. We call
$(e_M)_{M \in \Sub_s(S)}$ a \emph{collection} of decomposable elements, and denote it as
$(e_M)_M$.

\begin{definition}\label{def:IntScalingColl}
The \emph{core} of the collection $(e_M)_{M}$ is the subsystem $C \in \Sub(S)$ for which $\phi(e_C)$ is maximal.%
\footnote{If the maximum does not exist, we define the core to be the empty system $I$.}
The \emph{core integration scaling} of the collection is $\intscaling(e_C)$.
The \emph{core integration scaling} of a pair of collections $(e_M, f_M)_{M}$ is
$\intscaling(e_{C}, f_{D})$, where $C, D$ are the cores of $(e_M)_M$ and $(f_M)_M$, respectively.
\end{definition}

\section{Constructions - Mechanism Level} \label{sec:alg-mech}

Let $S \in \Sys$ be a physical system whose experience in a state $s \in \St(S)$ is to be determined.
The first level of the algorithm involves fixing some subsystem $M \in \Sub_s(S)$, referred to as a `mechanism', and associating to it an object called its `concept' which belongs to the \emph{concept space}
\begin{equation}\label{eq:Concepts}
\Concepts(S) := \PExp(S) \times \PExp(S) \:.
\end{equation}

For every choice of $P \in \Sub_s(S)$, called a `purview', the repertoire values $\caus_s(M,P)$ and $\eff_s(M,P)$ are elements of $\PExp(S)$ with given decompositions over $\mathbb{D}_{M,P}$. Fixing $M$, they form collection of decomposable elements,
\begin{align}\begin{split}\label{eq:CER}
\caus_s(M) &:= (\caus_s(M,P))_{P \in \Sub(S)} \\
\eff_s(M) &:= (\eff_s(M,P))_{P \in \Sub(S)} \:.
\end{split}\end{align}

The \emph{concept} of $M$ is then defined as the core integration scaling of this pair of collections,
\begin{equation}\begin{split}\label{eq:DefConcept}
\con{S,s}{M} := \textrm{Core integration scaling of }  \left(\caus_s(M), \eff_s(M) \right) \:.
\end{split}\end{equation}
It is an element of $\Concepts(S)$. Unravelling our definitions, the concept thus consists of the values of the cause and effect repertoires at their respective `core' purviews $P^c, P^e$, i.e.~those which make them `most integrated'. These values $\caus(M,P^c)$ and $\eff(M,P^e)$ are then each rescaled to have intensity given by the minima of their two integration levels.

\section{Constructions - System Level} \label{sec:alg-sys}

The second level of the algorithm specifies the experience of the system $S$ in state $s$.
To this end, all concepts of a system are collected to form its \emph{Q-shape}, defined as
\begin{equation} \label{eq:Q-Shape}
\QShape_s(S) := (\con{S,s}{M})_{M \in \Sub_s(S)}  \:.
\end{equation}
This is an element of the space 
\begin{equation}\label{eq:ExpS}
\Exp(S) = \Concepts(S)^{n(S)} \:,
\end{equation}
where $n(S) := |\Sub_s(S)|$, which is finite and independent of the state $s$ according to our assumptions.
We can also define a Q-shape for any cut of $S$. Let $z \in \mathbb D_S$ be a decomposition, $\cut{S}{z}$ the corresponding cut system and $\cut{s}{z}$ be the
corresponding cut state. We define
\begin{equation} \label{eq:Q-Shape-cuts}
\QShape_s(\cut{S}{z}) := (\con{\cut{S}{z},\cut{s}{z}}{M})_{M \in \Sub_{\cut{s}{z}}(\cut{S}{z})} \:.
\end{equation}
Because of~\eqref{eq:cestructure}, and since the number of subsystems remains the same when cutting, 
$\QShape_s(\cut{S}{z})$ is also an element of $\Exp(S)$.
This gives a map
\begin{align*}
\bar \QShape_{S,s}: \mathbb D_S &\rightarrow \Exp(S)\\
z &\mapsto \QShape_s(\cut{S}{z})
\end{align*}
which is a decomposition of $\QShape_s(S)$ over $\mathbb D_S$. Considering this map for every subsystem of $S$ gives a collection of decompositions
defined as
\begin{align*}
\QShape(S,s) := \big( \bar \QShape_{M,s|_M} \big)_{M \in \Sub_s(S)} \:
\end{align*}
This is the system level-object of relevance and is what specifies the experience of a system according to IIT.

\begin{definition}\label{def:ActualExp}
The actual \emph{experience} of the system $S$ in the state $s \in \St(S)$ is
\begin{equation}\label{eq:Exp}
\Exp(S,s) :=  \textrm{Core integration scaling of } \QShape(S,s) \:.
\end{equation}
\end{definition}
The definition implies that $\Exp(S,s) \in \Exp(M)$, where $M \in \Sub_s(S)$ is the core of the collection $\QShape(S,s)$, called the \emph{major complex}.
It describes which part of the system $S$ is actually conscious.
In most cases there will be a natural embedding $\Exp(M) \hookrightarrow \Exp(S)$ for a subsystem $M$ of $S$, allowing us to view $\Exp(S,s)$ as an element of $\Exp(S)$ itself. Assuming this embedding
to exist allows us to define an Integrated Information Theory concisely in the next section.

\section{Integrated Information Theories}\label{sec:IITs}

We can now summarise all that we have said about IITs.

\begin{definition}
An \emph{Integrated Information Theory} is determined as follows. The \emph{data} of the theory is a system class $\Sys$ along with a cause-effect structure. The theory then gives a mapping
\begin{equation} \label{eq:IIT-High-Level}
\begin{tikzcd}
\Sys\rar{\Exp} & \Expcat 
\end{tikzcd}
\end{equation}
into the class $\Expcat$ of all experience spaces, sending each system $S$ to its space of experiences $\Exp(S)$ defined in~\eqref{eq:ExpS}, and a mapping
\begin{align}\begin{split} \label{eq:IIT-On-States}
\St(S) &\to \Exp(S) \\ 
s &\mapsto \Exp(S,s) 
\end{split}\end{align}
which determines the experience of the system when in a state $s$, defined in~\eqref{eq:Exp}.

\noindent The \emph{quantity} of the system's experience is given by 
\begin{equation*}
\Phi(S,s) := \norm \Exp(S,s) \norm \:,
\end{equation*}
and the quality of the system's experience is given by the normalized experience $\hat \Exp(S,s)$.
The experience is located in the core of the collection $\QShape(S,s)$, called \emph{major complex}, which is a subsystem of $S$.
\end{definition}

In the next sections we specify the data of several example IITs.

%% file: 1_examples.tex
\section{Classical IIT}\label{sec:Classical}

In this section we show how IIT 3.0~\cite{mayner2018pyphi,marshall2016integrated,tononi.2015.scholarpedia,oizumi2014phenomenology} fits in into the framework 
developed here. A detailed explanation of how our earlier algorithm fits with the usual presentation of IIT is given in Appendix~\ref{sec:comparison}. In~\cite{IITProcess} we give an alternative categorical presentation of the theory. 

\subsection{Systems}\label{sec:ClassicalSysDef} We first describe the system class underlying classical IIT. Physical systems $S$ are considered to be built up of several components  $S_1, \dots, S_n$, called \emph{elements}.
Each element $S_i$ comes with a finite set of states $\St(S_i)$, equipped with a metric. A state of $S$ is given by specifying a state of each element, so that
\begin{equation}\label{eq:CSt}
\DetSt(S) = \DetSt(S_1) \times ... \times \DetSt(S_n) \:.
\end{equation}
We define a metric~$d$ on $\DetSt(S)$ by summing over the metrics of the element state spaces $\DetSt(S_i)$ and denote the collection of probability distributions over $\DetSt(S)$ by $\ProbSt(S)$. Note that we may view $\DetSt(S)$ as a subset of $\ProbSt(S)$ by identifying any $s \in \DetSt(S)$ with its Dirac distribution $\delta_s \in \ProbSt(S)$, which is why we abbreviate $\delta_s$ by $s$ occasionally in what follows.

Additionally, each system comes with a probabilistic (discrete) \emph{time evolution operator} or \emph{transition probability matrix}, sending each $s \in \DetSt(S)$ to a probabilistic state $T(s) \in \ProbSt(S)$. Equivalently it may be described as a convex-linear map 
\begin{equation}\label{eq:SysTPM}
T \colon \ProbSt(S) \to \ProbSt(S)  \:.
\end{equation}
Furthermore, the evolution $T$ is required to satisfy a property called \emph{conditional independence}, which we define shortly. 

The class $\Sys$ consists of all possible tuples $S= (\{S_i\}^n_{i=1}, T)$ of this kind, with the trivial system $I$ having only a single element with a single state and trivial time evolution.

\subsection{Conditioning and Marginalizing}
In what follows, we will need to consider two operations on the map $T$. Let $M$ be any subset of the elements of a system and $M^\Bot$ its complement. We again denote by $\DetSt(M)$ the Cartesian product of the states of all elements in $M$, and by
$\ProbSt(M)$ the probability distributions on~$\DetSt(M)$.
For any $p \in \ProbSt(M)$, we define the {\em conditioning}~\cite{mayner2018pyphi} of $T$ on $p$ as the map
\begin{align}\begin{split}\label{eq:Cond}
T\Cond{p} \colon \ProbSt(M^{\Bot}) \rightarrow \ProbSt(S) \\
p' \mapsto  T(p \cdot p') 
\end{split}\end{align}
where $p \cdot p'$ denotes the multiplication of these probability distributions to give a probability distribution over $S$. 
Next, the {\em marginal} of $T$ over $M$ is defined as the map 
\begin{align}\label{eq:marg-simple}
\Mar{M}T \colon \ProbSt(S) &\rightarrow  \ProbSt(M^\Bot)
\end{align}
such that for each $p \in \ProbSt(S)$ and $m_2 \in \St(M^\bot)$ we have
\begin{align}
\Mar{M}T(p)(m_2) &= \sum_{m_1 \in \St(M)} T(p)(m_1,m_2) \:.
\end{align}
In particular we write $T_i := \Mar{S_i^\bot} T$ for each $i=1,\dots,n$. Conditional independence of $T$ may now be defined as the requirement that 
\[
T(p) = \prod^n_{i=1} T_i(p) \qquad \textrm{ for all } p \in \Prob(S) \:,
\]
where the right-hand side is again a probability distribution over $\St(S)$.

\subsection{Subsystems, Decompositions and Cuts}\label{sec:ClassicalSysData}
Let a system $S$ in a state $s \in \St(S)$ be given. The subsystems are characterized by subsets of the elements that constitute~$S$.
For any subset $M = \{S_1 , ... , S_m\}$ of the elements of $S$, the corresponding subsystem is also
 denoted $M$ and $\St(M)$ is again given by the product of the $\St(S_i)$, with time evolution 
\begin{equation}\label{eq:SubsysTPM}
T_{M} :=\Mar{M^\Botl} T \Cond{s_{M^\Botl} } \:,
\end{equation}
where $s_{M^\Botl}$ is the restriction of the state $s$ to $\St(M)$ and $\Cond{s_{M^\Botl} }$ denotes the conditioning on the Dirac distribution $\delta_{s_{M^\Botl}}$.

The decomposition set $\mathbb D_S$ of a system $S$ consists of all partitions of the set $N$ of elements of $S$ into two disjoint sets $M$ and $M^\Bot$. We denote such a partition by $z = (M, M^\bot)$. The trivial decomposition $1$ is the pair $(N,\emptyset)$. 

For any decomposition $(M, M^\bot)$ the corresponding cut system $\cut{S}{(M,M^\bot)}$ is the same as $S$ but with a new time evolution $T^{(M,M^\bot)}$. Using conditional independence, it may be defined for each $i = 1, \dots, n$ as
\begin{equation}\label{eq:ClCut}
T^{(M,M^\bot)}_i
:=
\begin{cases}
T_i & i \in M^\bot \\
T_i\Cond{\uniform{M^\Botl}}\Mar{M^\Botl} & i \in M
\end{cases} 
\end{equation}
where $\uniform{M} \in \Prob(M)$ denotes the uniform distribution on $\St(M)$. 
This is interpreted in the graph depiction as removing all those edges from the graph whose source is in $M^\Bot$ and whose target is in $M$. The corresponding input of the target element is replaced by noise, i.e. the uniform probability distribution over the source element. 

\subsection{Proto-Experiences}
For each system $S$, the first Wasserstein metric (or `Earth Mover's Distance') makes $\Prob(S)$ a metric space $(\Prob(S),d)$.
The space of proto-experiences of classical IIT is
\begin{equation}\label{eq:ProtoClassical}
\PExp(S) := \overline{\ProbSt(S)} \:,
\end{equation}
where $\overline{\ProbSt(S)}$ is defined in Example~\ref{ex:IITProto}.
Thus elements of $\PExp(S)$ are of the form $(p, r)$ for some $p \in \Prob(S)$ and $r \in \R^+$, with distance function, intensity and scalar multiplication as defined in the example.

\subsection{Repertoires}\label{sec:CRep}
It remains to define the cause-effect repertoires. Fixing a state $s$ of $S$, the first step will be to define maps $\nonext{\caus}_s$ and $\nonext{\eff}_s$ which send any choice of $(M,P) \in \Sub(S) \times \Sub(S)$ to an element of $\ProbSt(P)$. These should describe the way in which the current state of $M$ constrains that of $P$ in the next or previous time-steps. We begin with the effect repertoire. For a single element purview $P_i$ we define 
\begin{equation}\label{eq:CauseEffPrep}
\nonext{\eff}_s(M,P_i) := \Mar{P_i^\Botl} T \Cond{ \uniform{M^\Botl}}(s_M) \:,
\end{equation}
where $s_M$ denotes (the Dirac distribution of) the restriction of the state $s$ to $M$.
While it is natural to use the same definition for arbitrary purviews, IIT 3.0 in fact uses another definition based on consideration of `virtual elements'~\cite{mayner2018pyphi,marshall2016integrated,tononi.2015.scholarpedia}, which also makes calculations more efficient~\cite[Supplement S1]{mayner2018pyphi}. For general purviews $P$, this definition is
\begin{equation}\label{eq:Virtual1}
\nonext{\eff}_s(M,P) = \prod_{P_i\in P} \nonext{\eff}_s(M,P_i) \:,
\end{equation}
taking the product over all elements $P_i$ in the purview $P$. Next, for the cause repertoire, for a single element mechanism $M_i$
and each $\tilde s \in \St(P)$, we define
\begin{equation} \label{eq:cause-v-simple}
\nonext{\caus}_s(M_i,P)[\tilde s]
=
\lambda \: \Mar{M_i^\bot} T \Cond{ \uniform{P^\Bot}}(\delta_{\tilde s})[s_{M_i}] \:,
\end{equation}
where $\lambda$ is the unique normalisation scalar making $\nonext{\caus}_s(M_i,P)$ a valid element of $\ProbSt(P)$. 
Here, for clarity, we have indicated evaluation of probability distributions at particular states by square brackets.
If the time evolution operator has an inverse $T^{-1}$, this cause repertoire could be defined similarly to~\eqref{eq:CauseEffPrep} by
$
\nonext{\caus}_s(M_i,P)
=  \Mar{P^\Bot} T^{-1} \Cond{ \uniform{M_i^\Bot}} (s_{M_i}) \:,
$
but classical IIT does not utilize this definition.

For general mechanisms $M$, we then define
\begin{equation} \label{eq:cause-full} 
\nonext{\caus}_s(M,P) = \kappa \prod_{M_i \in M} \nonext{\caus}_s(M_i,P)
\end{equation}
where the product is over all elements $M_i$ in $M$ and where $\kappa \in \R^+$ is again a normalisation constant. We may at last now define 
\begin{align}\begin{split}\label{eq:ClassicalCauseEffRep}
\caus_s(M,P) &:=  \nonext{\caus}_s(M,P) \cdot \nonext{\caus}_s(\emptyset,P^\Bot) 
 \\
\eff_s(M,P) &:=   \nonext{\eff}_s(M,P) \cdot \nonext{\eff}_s(\emptyset,P^\Bot) \:,
\end{split}
\end{align}
with intensity $1$ when viewed as elements of $\PExp(S)$.
Here, the dot indicates again the multiplication of probability distributions and $\emptyset$ denotes the empty mechanism.

The distributions $\nonext{\caus}_s(\emptyset,P^\Bot)$ and $\nonext{\eff}_s(\emptyset,P^\Bot)$ are called the \emph{unconstrained cause and effect repertoires} over~$P^\Bot$.

\begin{Remark}\label{rem:Vanishes}
It is in fact possible for the right-hand side of \eqref{eq:cause-v-simple} to be equal to $0$ for all $\tilde s$ for some $M_i \in M$. In this case we set $\caus_s(M,P) = (\uniform{S},0)$ in $\PExp(S)$.
\end{Remark}

Finally we must specify the decompositions of these elements over $\mathbb D_{M,P}$. For any partitions $z_M = (M_1,M_2)$ of $M$ and $z_P = (P_1,P_2) $ of $P$, we define 
\begin{align}\begin{split}\label{eq:CDecompDef}
&\overline{\caus_s}({M,P})(z_M, z_P) :=  \nonext{\caus}_s(M_1,P_1) \cdot \nonext{\caus}_s(M_2,P_2) \cdot \nonext{\caus}_s(\emptyset,P^\Bot)  \\
&\overline{\eff_s}({M,P})(z_M, z_P) :=  \nonext{\eff}_s(M_1,P_1) \cdot \nonext{\eff}_s(M_2^,P_2) \cdot \nonext{\eff}_s(\emptyset,P^\Bot) \:,
\end{split}\end{align}
where we have abused notation by equating each subset $M_1$ and $M_2$ of nodes with their induced subsystems of $S$ via the state $s$. 
\medskip

This concludes all data necessary to define classical IIT. If the generalized definition of Section~\ref{sec:IITs} is applied to this data, it yields precisely classical IIT 3.0 defined by Tononi et al. In Appendix~\ref{sec:comparison}, we explain in detail how our definition of IIT, equipped with this data, maps to the usual presentation of the theory.

\section{Quantum IIT} \label{sec:quantum-IIT}
In this section, we consider quantum IIT defined in~\cite{zanardi2018quantum}. This is also a special case of the definition in terms of process theories we give in~\cite{IITProcess}.

\subsection{Systems}\label{sec:QSys}
Similar to classical IIT, in quantum IIT systems are conceived as consisting of elements $\hilbH_1, ... \, , \hilbH_n$. Here, each element $\hilbH_i$ is described by a finite dimensional Hilbert space and the state space of the system $S$ is defined in terms of the element Hilbert spaces as
\[
\St(S) = \mathcal S(\H_S) \qquad \textrm{ with } \quad \H_S = \bigotimes_{i=1}^{n} \H_{i} \:,
\]
where $\mathcal S(\H_S) \subset L(\H_S)$ describes the positive semidefinite Hermitian operators of unit trace on $\H_S$, aka density matrices.
The time evolution of the system is again given by a time evolution operator, which here is assumed to be a trace preserving (and in \cite{zanardi2018quantum} typically unital) completely positive map
\[
\mathcal T: L(\H_S) \rightarrow L(\H_S) \:.
\]

\subsection{Subsystems, Decompositions and Cuts}\label{sec:QSysData}
Subsystems are again defined to consist of subsets $M$ of the elements of the system, with corresponding Hilbert space $\hilbH_M := \bigotimes_{i \in M} \hilbH_i$. The time-evolution $\mathcal T_M: L(\H_M) \rightarrow L(\H_M) $ is defined as
\[
 \mathcal T_M (\rho) = \tr_{M^\Botl} \big( \mathcal T ( \tr_{M^\Botl}(s) \otimes \rho ) \big) \:,
\]
where $s \in \mathcal S(\H_S)$ and $\tr_{M^\Botl}$ denotes the trace over the Hilbert space $\H_{M^\Botl}$.

Decompositions are also defined via partitions $z = (D,D^\Bot) \in \mathbb D_S$ of the set of elements~$N$ into two disjoint subsets $D$ and $D^\Bot$ whose union is $N$. For any such decomposition, the cut system $\cut{S}{(D,D^\Bot)}$ is defined to have the same set of states but time evolution
\[
\mathcal T^{(D,D^\Bot)} (s) = \mathcal T \big( \tr_{D^\Botl}(s) \otimes \, \discardflip{D^\Botl} \big) \:,
\]
where $\discardflip{D^\Botl}$  is the maximally  mixed state on $\hilbH_{D^\Bot}$, i.e.~$\discardflip{D^\Botl} = \frac{1}{\dim(\hilbH_{D^\Botl})} \, 1_{\H_{D^\Botl}}$.

\subsection{Proto-Experiences}\label{sec:QProtoExp}
For any $\rho, \sigma \in \mathcal S(\H_S)$, the trace distance defined as
\[
	d(\rho, \sigma) = \tfrac{1}{2} \, \tr_S \big( \sqrt{(\rho-\sigma)^2} \big) 
\]
turns $(S(\H_S), d)$ into a metric space. The space of proto-experiences is defined based on this metric space as described in Example~\ref{ex:IITProto},
\[
\PExp(S) := \overline{S(\H_S)} \:.
\]

\subsection{Repertoires}\label{sec:QRep}
We finally come to the definition of the cause- and effect repertoire. Unlike classical IIT, the definition in~\cite{zanardi2018quantum} does not consider virtual elements. 
Let a system $S$ in state $s \in \St(S)$ be given. As in Section~\ref{sec:CRep}, we utilize maps $\nonext{\caus}_s$ and $\nonext{\eff}_s$
which here map subsystems $M$ and $P$ to $\St(P)$. They are defined as
\begin{align*}
\nonext{\eff}_s(M,P) &= \tr_{P^\Botl} \mathcal T \big(  \tr_{M^\Botl} (s) \otimes \discardflip{M^\Botl} \big)\\
\nonext{\caus}_s(M,P) &= \tr_{P^\Botl} \mathcal T^\dagger \big(  \tr_{M^\Botl} (s) \otimes \discardflip{M^\Botl} \big)  \:,
\end{align*}
where $\mathcal T^\dagger$ is the Hermitian adjoint of $\mathcal T$. We then define
\begin{align*}
\caus_s(M,P) &:=  \nonext{\caus}_s(M,P) \otimes \nonext{\caus}_s(\emptyset,P^\Bot) \\
\eff(M,P)&:=   \nonext{\eff}_s(M,P) \otimes \nonext{\eff}_s(\emptyset,P^\Bot) \:,
\end{align*}
each with intensity $1$, where $\emptyset$ again denotes the empty mechanism.
Similarly, decompositions of these elements over $\mathbb D_{M,P}$ are defined as
\begin{align*}
&\overline{\caus_s}({M,P})(z_M, z_P) :=  \nonext{\caus}_s(M_1,P_1) \otimes \nonext{\caus}_s(M_2,P_2) \otimes \nonext{\caus}_s(\emptyset,P^\Bot) 
\\
&\overline{\eff_s}({M,P})(z_M, z_P) :=  \nonext{\eff}_s(M_1,P_1) \otimes \nonext{\eff}_s(M_2,P_2) \otimes \nonext{\eff}_s(\emptyset,P^\Bot)  \:,\end{align*}
again with intensity $1$, where $z_M = (M_1,M_2) \in \mathbb D_M$ and $z_P = (P_1,P_2) \in \mathbb D_P$.

%% file: 1_probs.tex
\section{Extensions of Classical IIT} \label{sec:problems}

The physical systems to which IIT 3.0 may be applied are limited in a number of ways: they must have a discrete time-evolution, satisfy Markovian dynamics and exhibit a discrete set of states~\cite{BarrettMediano.2019}. Since many physical systems do not satisfy these requirements, if IIT is to be taken as a fundamental theory about reality, it must be extended to overcome these limitations. 

In this section, we show how 
IIT can be redefined to cope with continuous time, non-Markovian dynamics and non-compact state spaces, by a redefinition of the maps~\eqref{eq:Virtual1} and~\eqref{eq:cause-full} and, in the case of non-compact state spaces, a slightly different choice of~\eqref{eq:ProtoClassical}, while leaving all of the remaining structure as it is. While we do not think that our particular definitions are satisfying as a general definition of IIT, these results show that the disentanglement of the essential mathematical structure of IIT from auxiliary tools (the particular definition of cause-effect repertoires used to date) can help to overcome fundamental mathematical or conceptual problems.

In Section~\ref{sec:Non-Canonical}, we also explain which solution to the problem of non-canonical metrics is suggested by our formalism. 

\subsection{Discrete Time and Markovian Dynamics}\label{sec:PTimeM}
In order to avoid the requirement of a discrete time and Markovian dynamics, instead of working with the time evolution
operator~\eqref{eq:SysTPM}, 
we define the cause- and effect repertoires in reference to a given trajectory of a physical state $s \in \St(S)$. 
The resulting definitions can be applied independently of whether trajectories are being determined by Markovian dynamics in a particular application, or not.\medskip

Let $t\in\I$ denote the time parameter of a physical system. If time is discrete, $\I$ is an ordered set. If time is continuous, $\I$ is an interval of reals. For simplicity, we assume $0 \in \I$. In the deterministic case, a trajectory of a state $s \in \St(S)$ is simply a curve in $\St(S)$, which we denote by $(s(t))_{t \in \I}$ with $s(0) = s$. 
For probabilistic systems (such as neural networks with a probabilistic update rule), it is a curve of probability distributions $\Prob(S)$, which we denote by $(p(t))_{t\in\I}$, with $p(0)$ equal to the Dirac distribution $\delta_s$. The latter case includes the former, again via Dirac distributions.

In what follows, we utilize the fact that in physics, state spaces are defined such that the dynamical laws of a system allow to determine the trajectory of each state.
Thus for every $s \in \St(S)$, there is a trajectory  $(p_s(t))_{t\in\I}$ which describes the time evolution of $s$.\medskip

The idea behind the following is to define, for every $M, P \in \Sub(S)$, a trajectory $p_{s}^{(P,M)}\!(t)$ in~$\Prob(P)$ which quantifies how much the state of the purview~$P$ at time $t$ is being constrained by imposing the state $s$ at time $t=0$ on the mechanism~$M$. 
This gives an alternative definition of the maps~\eqref{eq:Virtual1} and~\eqref{eq:cause-full}, while the rest of classical IIT can be applied as before.\medskip

Let now $M, P \in \Sub(S)$ and $s \in \St(S)$ be given.
We first consider the time evolution of the state $(s_M, v) \in \St(S)$, where $s_M$ denotes the restriction of $s$ to $\St(M)$ as before and where $v \in \St(M^\Bot)$ is an arbitrary state of $M^\Bot$. We denote the time evolution of this state by $p_{(s_M,v)}(t) \in \Prob(S)$.
Marginalizing this distribution over $P^\Bot$ gives a distribution on the states of $P$, which we denote as $ p_{(s_M,v)}^P(t) \in \Prob(P)$.
Finally, we average over $v$ using the uniform distribution $\uniform{M^\Botl}$. Because state spaces are finite in classical IIT, this averaging can be defined pointwise for every $w \in \St(P)$ by
\begin{equation}\label{eq:Average}
p_{s}^{(P,M)}(t)(w) := \kappa  \sum_{v \in \St(M^\Bot)} p_{(s_M,v)}^P(t)(w) \: \uniform{M^\Botl}(v) \: ,
\end{equation}
where $\kappa$ is the unique normalization constant which ensures that $p_{s}^{(P,M)}(t) \in \Prob(P)$.

The probability distribution $p_{s}^{(P,M)}(t) \in \Prob(P)$ describes how much the state of the purview~$P$ at time~$t$ is being constrained by imposing the state $s$ on $M$ at time $t=0$ as desired. Thus, for every $t \in \I$, we have obtained a mapping of two subsystems $M, P$ to an element $p_{s}^{(P,M)}(t)$ of $\Prob(P)$ which has the same interpretation as the map~\eqref{eq:CauseEffPrep} considered in classical IIT.
If deemed necessary, virtual elements could be introduced just as in~\eqref{eq:Virtual1} and~\eqref{eq:cause-full}.
\medskip
 	
So far, our construction can be applied for any time $t \in T$. It remains to fix this freedom in the choice of time.
For the discrete case, the obvious choice is to define~\eqref{eq:Virtual1} and~\eqref{eq:cause-full} in terms of neighbouring time-steps.
For the continuous case, several choices exist. E.g., one could consider the positive and negative semi-derivatives of 
$p_{s}^{(P,M)}(t)$ at $t=0$, in case they exist, or add an arbitrary but fixed time scale $\Delta$ to define
the cause- and effect repertoires in terms of $p_{s}^{(P,M)}(t_0\pm\Delta)$. However, the most reasonable choice is in our eyes to work with limits, in case they exist, by defining 
\begin{equation}
\nonext{\eff}_s(M,P) := \prod_{P_i\in P} \lim_{t \rightarrow \infty} p_{s}^{(P_i,M)}(t)
\end{equation}
to replace~\eqref{eq:Virtual1} and
\begin{equation}
\nonext{\caus}_s(M,P) := \kappa \  \prod_{M_i\in M} \lim_{t \rightarrow - \infty} p_{s}^{(P,M_i)}(t)
\end{equation}
to replace~\eqref{eq:cause-full}. The remainder of the definitions of classical IIT can then be applied as before.

\subsection{Discrete Set of States}
The problem with applying the definitions of classical IIT to systems with continuous state spaces (e.g. neuron membrane potentials~\cite{BarrettMediano.2019}) is that in certain cases, uniform probability distributions do not exist. E.g., if the state space of a system~$S$ consists of the positive real numbers~$\R^+$, no uniform distribution can be defined which has a finite total volume, so that no uniform \emph{probability} distribution $\uniform{S}$ exists.

It is important to note that this problem is less universal than one might think. E.g., if the state space of the system is a closed and bounded subset of $\R^+$, e.g. an interval $[a,b] \subset \R^+$, a uniform probability distribution can be defined using measure theory, which is in fact the natural mathematical language for probabilities and random variables. Nevertheless, the observation in~\cite{BarrettMediano.2019} is correct that if a system has a non-compact continuous state space, $\uniform{S}$ might not exist, which can be considered a problem w.r.t. the above-mentioned working hypothesis.

This  problem can be resolved for all well-understood physical systems by replacing the uniform probability distribution $\uniform{S}$ by some other mathematical entity which allows to define a notion of averaging states. An example is quantum theory (Section~\ref{sec:quantum-IIT}), whose state-spaces are continuous and non-compact. Here, the maximally mixed state $\discardflip{S}$ plays the role of the uniform probability distribution.
For all relevant classical systems with non-compact state spaces (whether continuous or not), the same is true: There exists a canonical uniform measure $\mu_S$ which allows to define the cause-effect repertoires similar to the last section, as we now explain. Examples for this canonical uniform measure are
the Lebesgue measure for subsets of $\R^n$~\cite{rudin2006real}, or the Haar measure for locally compact topological groups~\cite{salamon2016measure} such as Lie-groups.\medskip

In what follows, we explain how the construction of the last section needs to be modified in order to be applied to this case.
In all relevant classical physical theories, $\St(S)$ is a metric space in which every probability measure is a Radon measure, in particular locally finite,
and where a canonical locally finite uniform measure $\mu_S$ exists. We define $\Prob_1(S)$ to be the space of probability measures whose first moment is finite. For these, the first Wasserstein metric (or `Earth Mover's Distance') $W_1$ exists, so tat $(\Prob_1(S),W_1)$ is a metric space.

As before, the dynamical laws of the physical systems determine for every state $s \in \St(S)$ a time evolution $p_s(t)$, which here is an element of $\Prob_1(S)$. 
Integration of this probability measure over $\St(P^\Bot)$ yields the marginal probability measure $p_s^P(t)$.
As in the last section, we may consider these probability measures for the state $(s_M,v) \in \St(S)$, where $v \in \St(M^\Bot)$.
Since $\mu_S$ is not normalizable, we cannot define $p_{s}^{(P,M)}(t)$ as in~\eqref{eq:Average}, for the result might be infinite. 

Using the fact that $\mu_S$ is locally finite, we may, however, define a somewhat weaker equivalent. To this end, we note that
for every state $s_{M^\Botl}$, the local finiteness of $\mu_{M^\Botl}$ implies that there is a neighbourhood $N_{s,{M^\Botl}}$ in $\St(M^\Bot)$ for which $\mu_{M^\Botl}(N_{s,{M^\Botl}})$ is finite. We choose a sufficiently large neighbourhood which satisfies this condition. Assuming $p_{(s_M,v)}^P(t)$ to be a measurable function in $v$, for every $A$ in the $\sigma$-algebra of $\St(M^\Bot)$, we can thus  define
\begin{equation}\label{eq:Average2}
p_{s}^{(P,M)}(t)(A) := \kappa  \int_{N_{s,{M^\Botl}}} p_{(s_M,v)}^P(t)(A) \ d\mu_{M^\Botl}(v) \: ,
\end{equation}
which is a finite quantity. The $p_{s}^{(P,M)}(t)$ so defined is non-negative, vanishes for $A = \emptyset$ and satisfies countable additivity. Hence it is a measure on $\St(P)$ as desired, but might not be normalizable.

All that remains for this to give a cause-effect repertoire as in the last section, is to make sure that any measure (normalized or not) is an element of $\PExp(S)$. The theory is flexible enough to do this by setting $d(\mu,\nu)=| \mu - \nu | (\St(P))$ if either $\mu$ or $\nu$ is not in $\Prob_1(S)$, and $W_1(\mu,\nu)$ otherwise.
Here, $|\mu-\nu|$ denotes the total variation of the signed measure $\mu-\nu$, and $| \mu - \nu | (\St(P))$ is the volume thereof \cite{signedmeasure.encmath,halmos2013measure}. While not a metric space any more, the tuple $(\mathcal M(S), d)$, with $\mathcal M(S)$ denoting all measures on $\St(S)$, can still be turned into a space of proto-experiences as explained in Example~\ref{ex:IITProto}. This gives
\[
\PExp(S) := \overline{\mathcal M(S)} 
\]
and finally allows to construct cause-effect repertoires as in the last section.\medskip

\subsection{Non-Canonical Metrics}\label{sec:Non-Canonical}

Another criticism of IIT's mathematical structure mentioned \cite{BarrettMediano.2019} is that the metrics used in IIT's algorithm are, to a certain extend, chosen arbitrarily. Different choices indeed imply different results of the algorithm, both concerning the quantity and quality of experience, which can be considered problematic.

The resolution of this problem is, however, not so much a technical as a conceptual or philosophical task, for what is needed to resolve this issue is a justification of why a particular metric should be used. Various justifications are conceivable, e.g. identification of desired behaviour of the algorithm when applied to simple systems. When considering our mathematical reconstruction of the theory, the following natural justification offers itself.

Implicit in our definition of the theory as a map from systems to experience spaces is the idea that the mathematical structure of experiences spaces (Definition~\ref{def:exp-space}) reflects the phenomenological structure of experience. This is so, most crucially, for the distance function $d$, which describes how similar two elements of experience spaces are. Since every element of an experience space corresponds to a conscious experience, it is naturally to demand that the similarly of the two mathematical objects should reflect the similarity of the experiences they describe. Put differently, the distance function $d$ of an experience space should in fact mirror (or ``model'') the similarity of conscious experiences as experienced by an experiencing subject.

This suggests that the metrics $d$ used in the IIT algorithm should, ultimately, be defined in terms of the phenomenological structure of similarity of conscious experiences. For the case of colour qualia, this is in fact feasible \cite[Example\,3.18]{kleiner2019models}, \cite{Kuehni.2010,SharmaWuDalal.2004}.
In general, the mathematical structure of experience spaces should be intimately tied to the phenomenology of experience, in our eyes.

%% file: 1_discussion.tex
\section{Summary \& Outlook} \label{sec:discussion}

In this article, we have propounded the mathematical structure of Integrated Information Theory.
First, we have studied which exact structures the IIT algorithm uses in the mathematical description of physical systems, on the one hand, and in the mathematical description of conscious experience, on the other. Our findings are the basis of definitions of a physical system class $\Sys$ and a class $\Expcat$ of experience spaces, and allowed us to view IIT as a map $\Sys \rightarrow \Expcat$.

Next, we needed to disentangle the essential mathematics of the theory from auxiliary formal tools used in the contemporary definition. To this end, we have introduced the precise notion of decomposition of elements of an experience space required by the IIT algorithm. The pivotal cause-effect repertoires are examples of decompositions so defined, which allowed us to view any particular choice, e.g. the one of `classical' IIT developed by Tononi et. al., or the one of `quantum' IIT recently introduced by Zanardi et. al. as data provided to a general IIT algorithm.

The formalization of cause-effect repertoires in terms of decompositions then led us to define the essential ingredients of IIT's algorithm concisely in terms of integration levels, integration scalings and cores. These definitions describe and unify recurrent mathematical operations in the contemporary presentation, and finally allowed to define IIT completely in terms of a few lines of definition.

Throughout the paper, we have taken great care to make sure our definitions reproduce exactly the contemporary version of IIT 3.0. The result of our work is a mathematically rigorous and general definition of Integrated Information Theory. This definition can be applied to any meaningful notion of systems and cause-effect repertoires, and we have shown that this allows to overcome most of the mathematical problems of the contemporary definition identified to date in the literature.

We believe that our mathematical reconstruction of the theory can be the basis for refined mathematical and philosophical analysis of IIT. We also hope that this mathematisation may make the theory more amenable to study by mathematicians, physicists, computer scientists and other researchers with a strongly formal background.

\subsection{Process Theories}
Our generalization of IIT is axiomatic in the sense that we have only included those formal structures in the definition which are necessary for the IIT algorithm to be applied. This ensured that our reconstruction is as general as possible, while still true to IIT 3.0. As a result, several notions used in classical IIT, e.g., system decomposition, subsystems or causation, are merely defined abstractly at first, without any reference to the usual interpretation of these concepts in physics.

In the related article~\cite{IITProcess}, we show that these concepts can be meaningfully defined in any suitable \emph{process theory} of physics, formulated in the language of \emph{symmetric monoidal categories}. This approach can describe both classical and quantum IIT and yields a complete formulation of contemporary IIT in a categorical framework. 

\subsection{Further Development of IIT}
IIT is constantly under development, with new and refined definitions being added every few years. We hope that our mathematical analysis of the theory might help to contribute to this development. E.g., the working hypothesis that IIT is a fundamental theory, i.e. describes reality as it is, implies that technical problems of the theory need to be resolved. We have shown that our formalization allows address the technical problems mentioned in the literature. However, there are others which we have not addressed in this paper.

Most crucially, the IIT algorithm uses a series of maximalization and minimalization operations, unified in the notion of \emph{core} subsystems in our formalization. In general, there is no guarantee that these operations lead to unique results, neither in classical nor quantum IIT. Using different cores has major impact on the output of the algorithm, including the $\Phi$ value, which is a case of ill-definedness.\footnote{The problem of `unique existence' has been studied extensively in category theory using \emph{universal properties} and the notion of a \emph{limit}. Rather than requiring that each $E \in \Exp$ come with a metric, it may be possible to alter the IIT algorithm into a well-defined categorical form involving limits to resolve this problem.}

Furthermore, the contemporary definition of IIT as well as our formalization rely on there being a finite number of subsystems of each system, which might not be the case in reality. Our formalisation may be extendable to the infinite case by assuming that every system has a fixed but potentially infinite indexing set 
$\Sub(S)$, so that each $\Sub_s(S)$ is the image of a mapping $\Sub(S) \times \St(S) \to \Sys$, but we have not considered this in detail in this paper.

Finally, concerning more operational questions, it would be desirably to develop the connection to empirical measures such as the Perturbational Complexity Index PCI \cite{casarotto2016stratification,casali2013theoretically} in more detail, as well as to define a controlled approximation of the theory whose calculation is less expensive. Both of these tasks may be achievable by substituting parts of our formalization with simpler mathematical structure.

On the conceptual side of things, it would be desirable to have a more proper understanding of how the mathematical structure of experiences spaces corresponds to the phenomenology of experience, both for the general definition used in our formalization and the specific definitions used in classical and quantum IIT. In particular, it would be desirable to understand how it relates to the important notion of qualia, which is often asserted to have characteristic features such as ineffability, intrinsicality, non-contextuality, transparency or homogeneity~\cite{metzinger2006grundkurs}. For a first analysis towards this goal, cf.~\cite{kleiner2019models}.

%% file: acknowledgements.tex
\medskip

\Thanks {\small{\textsc{Acknowledgements:}}
We would like to thank the organizers and participants of the \emph{Workshop on Information Theory and Consciousness}
at the Centre for Mathematical Sciences of the University of Cambridge, of the \emph{Modelling Consciousness Workshop} in Dorfgastein
and of the \emph{Models of Consciousness Conference} at the Mathematical Institute of the University of Oxford for discussions on this topic.
Much of this work was carried out while Sean Tull was under the support of an EPSRC Doctoral Prize at the University of Oxford, from November 2018 to July 2019, and while Johannes Kleiner was under the support of postdoctoral funding at the Institute for Theoretical Physics of the Leibniz University of Hanover. We would like to thank both institutions.
}

%% file: 1_appendix.tex
\appendix

\section{Comparison with standard presentation of IIT 3.0} \label{sec:comparison}
In Section~\ref{sec:Classical}, we have defined the system class and cause-effect repertoires which underlie classical IIT.
The goal of this appendix is to explain in detail why applying our definition of the IIT algorithm yields IIT 3.0 defined by Tononi et al. In doing so, we will mainly refer to the terminology used in~\cite{tononi.2015.scholarpedia}, \cite{mayner2018pyphi}, \cite{oizumi2014phenomenology} and~\cite{marshall2016integrated}. We remark that a particularly detailed presentation of the algorithm of the theory, and of how the cause and effect repertoire are calculated, is given in the supplementary material S1 of \cite{mayner2018pyphi}.

\subsection{Physical Systems}
The systems of classical IIT are given in Section~\ref{sec:ClassicalSysDef}. They are often represented as graphs whose nodes are the elements $S_1, \dots , S_n$ and edges represent functional dependence, thus describing the time evolution of the system as a whole, which we have taken as primitive in~\eqref{eq:SysTPM}. This is similar to the presentation of the theory in terms of a transition probability function
\[ p: \DetSt(S) \times \DetSt(S) \rightarrow [0,1]
\]
in~\cite{marshall2016integrated}. For each probability distribution $\tilde p$ over $\DetSt(S)$, this relates to our time evolution operator $T$ via 
\[
T( \tilde p)[v] := \sum_{w \in \DetSt(S)} p(v,w) \ \tilde p(w) \:.
\]

\subsection{Cause-Effect Repertoires}
In contemporary presentations of the theory (\cite[p.\,14]{marshall2016integrated} or \cite{tononi.2015.scholarpedia}), the effect repertoire
is defined as 
\begin{equation}\label{eq:IITEffR}
p_{\textrm{effect}}(z_i, m_t) := \frac{1}{|\Omega_{M^c}|} \sum_{m^c \in \Omega_{M^c}} p \big( \, z_i \, | \, \textrm{do}(m_t, m^c) \, \big) \qquad z_i \in \Omega_{Z_i} 
\end{equation}
and
\begin{equation}\label{eq:IITEffR2}
p_{\textrm{effect}}(z, m_t) := \prod_{i=1}^{|z|} \ p_{\textrm{effect}}(z_i, m_t) \:.
\end{equation}
Here, $m_t$ denotes a state of the mechanism $M$ at time $t$. $M^c$ denotes the complement of the mechanism, denoted in our case as $M^\Bot$, $\Omega_{M^c}$ denotes the state space of the complement, and $m^c$ an element thereof. 
$Z_i$ denotes an element of the purview $Z$ (designated by $P$ in our case), $\Omega_{Z_i}$ denotes the state space of this element, $z_i$ a state of this element and $z$ a state of the whole purview. $|\Omega_{M^c}|$ denotes the cardinality of the state space of $M^c$, and $|z|$ equals the number of elements in the purview.
Finally, the expression $\textrm{do}(m_t, m^c)$ denotes a variant of the so-called ``do-operator''. It indicates that the state of the system, here at time $t$, is to be set to the term in brackets. This is called \emph{perturbing the system} into the state $(m_t, m^c)$. The notation $p ( z_i |  \textrm{do}(m_t, m^c) )$ then gives the probability of finding the purview element in the state $z_i$ at time $t+1$ given that the system is prepared in the state $(m_t, m^c)$ at time $t$.

In our notation, the right hand side of~\eqref{eq:IITEffR} is exactly given by the right-hand side of \eqref{eq:CauseEffPrep}, i.e. $\nonext{\eff}_s(M,P_i)$. The system is prepared in a uniform distribution on $M^c$ (described by the sum and prefactor in~\eqref{eq:IITEffR}) and with the restriction $s_M$ of the system state, here denoted by $m_t$, on $M$. Subsequently, $T$ is applied to evolve the system to time $t+1$, and the marginalization $\Mar{P_i^\Bot}$ throws away all parts of the states except those of the purview element $P_i$ (denoted above as $Z_i$). In total,~\eqref{eq:CauseEffPrep} is a probability distribution on the states of the purview element. When evaluating this probability distribution at one particular state $z_i$ of the element, one obtains the same numerical value as~\eqref{eq:IITEffR}. Finally, taking the product in \eqref{eq:IITEffR2} corresponds exactly to taking the product in~\eqref{eq:Virtual1}.\medskip

Similarly, the cause repertoire is defined as (\cite[p.\,14]{marshall2016integrated} or \cite{tononi.2015.scholarpedia})
\begin{equation}\label{eq:IITCauseR}
p_{\textrm{cause}} (z | m_{i,t} ) := \frac{\sum_{z^c \in \Omega_{Z^c}} p\big(\, m_{i,t} \, | \, \textrm{do}(z,z^c) \,  \big) }{\sum_{s \in \Omega_S} p\big(\,m_{i,t} \, | \, \textrm{do}(s) \, \big)} \qquad z \in \Omega_{Z_{t-1}} 
\end{equation}
and
\begin{equation}\label{eq:IITCauseR2}
p_{\textrm{cause}} (z | m_t ) := \frac{1}{K} \, \prod_{i=1}^{|m_t|} \ p_{\textrm{cause}} (z | m_{i,t} ) \:,
\end{equation}
where $m_i$ denotes the state of one element of the mechanism $M$, with the subscript~$t$ indicating that the state is considered at time $t$. $Z$ again denotes a purview, $z$ is a state of the purview and $\Omega_{Z_{t-1}}$ denotes the state space of the purview, where the subscript indicates that the state is considered at time $t-1$. $K$ denotes a normalization constant and $|m_t|$ gives the number of elements in $M$.

Here, the whole right hand side of~\eqref{eq:IITCauseR} gives the probability of finding the purview in state $z$ at time $t-1$ if the system is prepared in state $m_{i,t}$ at time $t$. In our terminology this same distribution is given by \eqref{eq:cause-v-simple},
where $\lambda$ is the denominator in~\eqref{eq:IITCauseR}. Taking the product of these distributions and re-normalising is then precisely \eqref{eq:cause-full}.

As a result, the cause and effect repertoire in the sense of \cite{oizumi2014phenomenology} correspond precisely in our notation to $\nonext{\caus}_s(M,P)$ and $\nonext{\eff}_s(M,P)$, each being distributions over $\St(P)$.
In~\cite[S1]{mayner2018pyphi}, it is explained that these need to be extended by the unconstrained repertoires before being used in the IIT algorithm, which in our formalization is done in~\eqref{eq:ClassicalCauseEffRep}, so that the cause-effect repertoires are now distributions over $\St(S)$.
These are in fact precisely what are called the \emph{extended} cause and effect repertoires or \emph{expansion to full state space} of the repertoires in \cite{oizumi2014phenomenology}.

The behaviour of the cause- and effect-repertoires when decomposing a system is described, in our formalism, by decompositions (Definition~\ref{def:Decomp}). Hence a decomposition $z \in \mathbb D_S$ is what is called a \emph{parition} in the classical formalism. For the case of classical IIT, a decomposition is given precisely by a partition of the set of elements of a system, and the cause-effect repertoires belonging to the decomposition are defined in~\eqref{eq:CDecompDef}, which corresponds exactly to the definition
\[
p_{\textrm{cause}}^{\textrm{cut}} (z | m_t) = p_{\textrm{cause}}(z^{(1)} | m_t^{(1)}) \times  p_{\textrm{cause}}(z^{(2)} | m_t^{(2)})
\]
in~\cite{marshall2016integrated}, when expanded to the full state space, and equally so for the effect repertoire.

\subsection{Algorithm - Mechanism Level}

Next, we explicitly unpack our form of the IIT algorithm to see how it compares in the case of classical IIT with \cite{oizumi2014phenomenology}.
In our formalism, the integrated information $\varphi$ of a mechanism $M$ of system $S$ when in state $s$ is
\begin{equation}\label{eq:APhimax}
\phimax(M) = \norm \con{S,s}{M}  \norm
\end{equation}
defined in Equation~\eqref{eq:DefConcept}. This definition conjoins several steps in the definition of classical IIT. To explain why it corresponds exactly to classical IIT, we disentangle this definition step by step.

First, consider $\caus_s(M,P)$ in Equation~\eqref{eq:CER}. This is, by definition, a decomposition map. The calculation of the integration level of this decomposition map, cf. Equation~\eqref{eq:IntegrLevel}, amounts to comparing $\caus_s(M,P)$ to the cause-effect repertoire associated with every decomposition using the metric of the target space $\PExp(S)$, which for classical IIT is defined in~\eqref{eq:ProtoClassical} and Example~\ref{ex:IITProto}, so that the metric $d$
used for comparison is indeed the Earth Mover's Distance. Since cause-effect repertoires have, by definition, unit intensity, the factor $r$ in the definition~\eqref{eq:ExIITProto} of the metric does not play a role at this stage. Therefore, the integration level of $\caus_s(M,P)$ is exactly the \emph{integrated cause information}, denoted as
\[
	\varphi_{\textrm{cause}}^{\textrm{MIP}}( y_t, Z_{t-1})
\]
in~\cite{tononi.2015.scholarpedia}, where $y_t$ denotes the (induced state of the) mechanism $M$ in this notation, and $Z_{t-1}$ denotes the purview $P$.  Similarly, the integration level of
$\eff_s(M,P)$ is exactly the \emph{integrated effect information}, denoted as 
\[
\varphi_{\textrm{effect}}^{\textrm{MIP}}( y_t, Z_{t+1}) \:.
\]

The integration scaling in~\eqref{eq:DefConcept} simply changes the intensity of an element of $\PExp(S)$ to match the integration level, using the scalar multiplication, which is important for the system level definitions. When applied to $\caus_s(M,P)$, this would result in an element of $\PExp(S)$ whose intensity is precisely $\varphi_{\textrm{cause}}^{\textrm{MIP}}( y_t, Z_{t-1})$.

Consider now the collections~\eqref{eq:CER} of decomposition maps. Applying Definition~\ref{def:IntScalingColl}, the core of $\caus_s(M)$ is that purview $P$ which gives the decomposition $\caus_s(M,P)$ with the highest integration level, i.e. with the highest $\varphi_{\textrm{cause}}^{\textrm{MIP}}( y_t, Z_{t-1})$. This is called the \emph{core cause}~$P^c$ of $M$, and similarly the core of $\eff_s(M)$ is called the \emph{core effect}~$P^e$ of $M$.

Finally, to fully account for~\eqref{eq:DefConcept}, we note that the integration scaling of a pair of decomposition maps rescales both elements to the minimum of the two integration levels. Hence the integration scaling of the pair $(\caus_s(M,P),\eff(M,P'))$ fixes the scalar value of both elements to be exactly the \emph{integrated information}, denoted as
\[
\varphi(y_t,Z_{t\pm1}) = \min \big(\varphi_{\textrm{cause}}^{\textrm{MIP}},\varphi_{\textrm{effect}}^{\textrm{MIP}} \big)
\]
in~\cite{tononi.2015.scholarpedia}, where $P = Z_{t+1}$ and $P'=Z_{t-1}$.

In summary, the following operations are combined in Equation~\eqref{eq:DefConcept}. The core of $ \left(\caus_s(M), \eff_s(M) \right)$ picks out the core cause $P^c$ and core effect $P^e$. The core integration scaling subsequently considers the pair $(\caus_s(M,P^c),\eff(M,P^e))$, called \emph{maximally irreducible cause-effect repertoire}, and determines the integration level of each by analysing the behaviour with respect to decompositions. Finally, it rescales both to the minimum of the integration levels. Thus it gives exactly what is called $\varphi^{\textrm{max}}$ in~\cite{tononi.2015.scholarpedia}. Using, finally, the definition of the intensity of the product $\PExp(S) \times \PExp(S)$ in Definition~\ref{def:prod-spaces}, this implies~\eqref{eq:APhimax}.
The concept $M$ in our formalization is given by the tuple 
\[
\con{S,s}{M}
:=
\big((\caus_s(M,P^c), \varphi^{\textrm{max}}(M)), (\eff_s(M,P^e),  \varphi^{\textrm{max}}(M))\big)
\]
i.e. the pair of maximally irreducible repertoires scaled by $\varphi^{\textrm{max}}(M)$. This is equivalent to
what is called a \emph{concept}, or sometimes \emph{quale sensu stricto}, in classcial IIT~\cite{tononi.2015.scholarpedia}, and denoted as $q(y_t)$.

We finally remark that it is also possible in classical IIT that a cause repertoire value $\caus_s(M,P)$ vanishes (Remark~\ref{rem:Vanishes}). In our formalization, it would hence be represented by $(\uniform{S},0)$ in $\PExp(S)$, so that $d(\caus_s(M,P),q) = 0$ for all $q \in \Exp(S)$ according to~\eqref{eq:ExIITProto}, which certainly ensures that $\varphi_{\textrm{cause}}^{\textrm{MIP}}(M,P) = 0$.

\subsection{Algorithm - System Level}
We finally explain how the system level definitions correspond to the usual definition of classical IIT.

The Q-shape $\QShape_s(S)$ is the collection of all concepts specified by the mechanisms of a system. Since each concept has intensity given by the corresponding integrated information of the mechanism, this makes $\QShape_s(S)$ what is usually called the \emph{conceptual structure} or \emph{cause-effect structure}. 
In \cite{oizumi2014phenomenology}, one does not include a concept for any mechanism $M$ with $ \varphi^{\textrm{max}}(M) = 0$. This manual exclusion is unnecessary in our case because the mathematical structure of experience spaces implies that mechanisms with $\phimax(M) = 0$ should be interpreted as having no conscious experience, and the algorithm in fact implies that they have `no effect'. Indeed we will now see that they do not contribute to the distances in $\Exp(S)$ or any $\Phi$ values, and so we do not manually exclude them.

When comparing $\QShape_s(S)$ with the Q-shape~\eqref{eq:Q-Shape-cuts} obtained after replacing $S$ by any of its cuts, it is important to note that both are elements of $\Exp(S)$ defined in~\eqref{eq:ExpS}, which is a product of experience spaces. According to Definition~\ref{def:prod-spaces}, the distance function on this product is
\[
d(\QShape_s(S),\QShape_s(\cut{S}{z}))
:= 
\sum_{M \in \Sub(S)}
d(\con{S,s}{M},\con{\cut{S}{z},\cut{s}{z}}{M}) \:.
\]
Using Definition \ref{ex:IITProto} and the fact that each concept's intensity is $\phimax(M)$ according to the mechanism level definitions,
each distance $d(\con{S,s}{M},\con{\cut{S}{z},\cut{s}{z}}{M})$  is equal to
\begin{align}\begin{split}\label{eq:AppDist}
\phimax(M) \cdot 
\big(d\big(&\caus_s(M,P^c_M), 
\caus^z_s(M,P^{z,c}_M)\big) \\
&+ 
d\big(\eff_s(M,P^e_M), 
\eff^z_s(M,P^{z,e}_M)\big)
\big) \: ,
\end{split}\end{align}
where $\phimax(M)$ denotes the integrated information of the concept in the original system $S$, and where the right-hand cause and effect repertoires are those of $\cut{S}{z}$ at its own core causes and effects for $M$. 
The factor $\phimax(M)$ ensures that the distance used here corresponds precisely to the distance used in \cite{oizumi2014phenomenology}, there called the \emph{extended Earth Mover's Distance}. If the integrated information $\phimax(M)$ of a mechanism is non-zero, it follows that $d(\con{S,s}{M},\con{\cut{S}{z},\cut{s}{z}}{M}) = 0$ as mentioned above, so that this concept does not contribute.

We remark that in~\cite[S1]{mayner2018pyphi}, an additional step is mentioned which is not described in any of the other papers we consider. Namely, if the integrated information of a mechanism is non-zero before cutting but zero after cutting, what is compared is not the distance of the corresponding concepts as in~\eqref{eq:AppDist}, but in fact the distance of the original concept with a special null concept, defined to be the unconstrained repertoire of the cut system. We have not included this step in our definitions, but it could be included by adding a choice of distinguished points to Example~\ref{ex:IITProto} and redefining the metric correspondingly.

In Equation~\eqref{eq:Exp} the above comparison is being conducted for every subsystem of a system $S$.
The subsystems of $S$ are what is called \emph{candidate systems} in~\cite{oizumi2014phenomenology}, and which describe that `part' of the system that is going to be conscious according to the theory (cf. below).  Crucially, candidate systems are subsystems of $S$, whose time evolution is defined in~\eqref{eq:SubsysTPM}. This definition ensures that the state of the elements of $S$ which are not part of the candidate system are fixed in their current state, i.e. constitute \emph{background conditions} as required in the contemporary version of classcial IIT~\cite{mayner2018pyphi}.

Equation~\eqref{eq:Exp} then compares the Q-shape of every candidate system to the Q-shape of all of its cuts, using the distance function described above, where the cuts are defined in~\eqref{eq:ClCut}. The cut system with the smallest distance gives the system-level \emph{minimum information partition} and the \emph{integrated (conceptual) information} of that candidate system, denoted as $\Phi(x_t)$ in~\cite{tononi.2015.scholarpedia}. 

The core integration scaling finally picks out that candidate system with the largest integrated information value. This candidate system is the \emph{major complex} $M$ of $S$, the part of $S$ which is conscious according to the theory as part of the \emph{exclusion postulate} of IIT. Its Q-shape is the 
\emph{maximally irreducible conceptual structure (MICS)}, also called \emph{quale sensu lato}. The overall \emph{integrated conceptual information} is, finally, simply the intensity of $\Exp(S,s)$ as defined in~\eqref{eq:Exp},
\[
\Phi(S,s) = \Exp(S,s) \:.
\]

\subsection{Constellation in Qualia Space}

Expanding our definitions, and denoting the major complex by $M$ with state $m=s|_M$,
in our terminology the actual experience of the system $S$ state $s$ is 
\begin{equation} \label{eq:Exp-explicit}
\Exp(S,s) := \, \frac{\Phi(M,m)}{ \norm \QShape_{m}(M) \norm } \cdot \, \QShape_{m}(M) \:.
\end{equation}
This encodes the Q-shape $\QShape_{m}(M)$, i.e. the maximally irreducible conceptual structure of the major complex, sometimes called \emph{quale sensu lato}, which is taken to describe the quality of conscious experience. By construction it also encodes the integrated conceptual information of the major complex, which captures its intensity, since we have $\norm \Exp(S,s)\norm = \Phi(M,m)$. The rescaling of $\QShape_{m}(M)$ in \eqref{eq:Exp-explicit} leaves the relative intensities of the concepts in the MICS intact. Thus $\Exp(S,s)$ is the \emph{constellation of concepts in qualia space} $\Exp(M)$ of~\cite{oizumi2014phenomenology}.

\vfill